\documentclass[%
reprint,
superscriptaddress,
 amsmath,amssymb,
 aps,
pra,
floatfix,
]{revtex4-2}
\usepackage{dcolumn}
\usepackage{bm}
\usepackage[utf8]{inputenc}
\usepackage[a4paper]{geometry}
\usepackage{fancyhdr}
\usepackage{graphicx}
\usepackage{dcolumn}
\usepackage{subfigure}
\usepackage{float}
\usepackage{schemata}
\usepackage{rotating}
\usepackage[table,xcdraw]{xcolor}
\usepackage{color}
\usepackage{amsmath}
\usepackage{booktabs}
 \usepackage{ulem}
\usepackage{amssymb}
\usepackage{multirow}

\providecommand{\keywords}[1]
{
  \small	
  \textbf{\textit{Keywords---}} #1
}
\begin{document}

\preprint{APS/123-QED}

\title{Self-induced spin pumping and inverse spin Hall effect in single FePt thin films}

\author{J. L. Ampuero}%
\email{jose.ampuero@ib.edu.ar}
\affiliation{Instituto de Nanociencia y Nanotecnología (CNEA - CONICET), Nodo Bariloche, Av. Bustillo 9500, (8400) Bariloche (RN), Argentina}
\affiliation{Laboratorio Resonancias Magnéticas, Centro Atómico Bariloche, Av. Bustillo 9500, (8400) Bariloche (RN), Argentina}
\affiliation{Instituto Balseiro, Universidad Nacional de Cuyo, Av. Bustillo 9500, 8400, San Carlos de Bariloche, Río Negro, Argentina}

\author{A. Anadón}
\affiliation{Institut Jean Lamour, Université de Lorraine / CNRS, UMR7198, 54011 Nancy, France}

\author{H. Damas}
\affiliation{Institut Jean Lamour, Université de Lorraine / CNRS, UMR7198, 54011 Nancy, France}

\author{J. Ghanbaja}
\affiliation{Institut Jean Lamour, Université de Lorraine / CNRS, UMR7198, 54011 Nancy, France}

\author{S. Petit-Watelot}
\affiliation{Institut Jean Lamour, Université de Lorraine / CNRS, UMR7198, 54011 Nancy, France}

\author{J.-C. Rojas-Sánchez}
\email{juan-carlos.rojas-sanchez@univ-lorraine.fr}
\affiliation{Institut Jean Lamour, Université de Lorraine / CNRS, UMR7198, 54011 Nancy, France}

\author{D. Velázquez Rodriguez}
\affiliation{Instituto de Nanociencia y Nanotecnología (CNEA - CONICET), Nodo Bariloche, Av. Bustillo 9500, (8400) Bariloche (RN), Argentina}
\affiliation{Laboratorio Resonancias Magnéticas, Centro Atómico Bariloche, Av. Bustillo 9500, (8400) Bariloche (RN), Argentina}
\affiliation{Instituto Balseiro, Universidad Nacional de Cuyo, Av. Bustillo 9500, 8400, San Carlos de Bariloche, Río Negro, Argentina}

\author{J. E. Gómez}
\affiliation{Instituto de Nanociencia y Nanotecnología (CNEA - CONICET), Nodo Bariloche, Av. Bustillo 9500, (8400) Bariloche (RN), Argentina}
\affiliation{Laboratorio Resonancias Magnéticas, Centro Atómico Bariloche, Av. Bustillo 9500, (8400) Bariloche (RN), Argentina}

\author{A. Butera}
\email{butera@cab.cnea.gov.ar}
\affiliation{Instituto de Nanociencia y Nanotecnología (CNEA - CONICET), Nodo Bariloche, Av. Bustillo 9500, (8400) Bariloche (RN), Argentina}
\affiliation{Laboratorio Resonancias Magnéticas, Centro Atómico Bariloche, Av. Bustillo 9500, (8400) Bariloche (RN), Argentina}
\affiliation{Instituto Balseiro, Universidad Nacional de Cuyo, Av. Bustillo 9500, 8400, San Carlos de Bariloche, Río Negro, Argentina}

\author{L. Avilés-Félix}
\affiliation{Instituto de Nanociencia y Nanotecnología (CNEA - CONICET), Nodo Bariloche, Av. Bustillo 9500, (8400) Bariloche (RN), Argentina}
\affiliation{Laboratorio Resonancias Magnéticas, Centro Atómico Bariloche, Av. Bustillo 9500, (8400) Bariloche (RN), Argentina}
\affiliation{Instituto Balseiro, Universidad Nacional de Cuyo, Av. Bustillo 9500, 8400, San Carlos de Bariloche, Río Negro, Argentina}

\raggedbottom

\begin{abstract} 
Self-induced spin Hall effect and self-torque hold great promise in the field of spintronics, offering a path toward highly efficient spin-to-charge interconversion, a pivotal advancement for data storage, sensing devices, or unconventional computing.  In this study, we investigate the spin-charge current conversion characteristics of chemically disordered ferromagnetic single FePt thin films by spin-pumping ferromagnetic resonance experiments performed on both a resonance cavity and on patterned devices. We clearly observe a self-induced signal in a single FePt layer. The sign of a single FePt spin pumping voltage signal is consistent with a typical bilayer with a positive spin Hall angle layer such as that of Pt on top of a ferromagnet (FM), substrate//FM/Pt. Structural analysis shows a strong composition gradient due to natural oxidation at both FePt interfaces, with the Si substrate and with the air. The FePt-thickness dependence of the self-induced charge current produced allowed us to obtain $\lambda _ \text{FePt}=(1.5\pm 0.1)$ nm and self-induced $\theta_ \text{self-FePt}=0.047 \pm 0.003$, with efficiency for reciprocal effects applications $\theta _ \text{self-FePt} \times \lambda _ \text{FePt} = 0.071$ nm which is comparable to that of Pt, $\theta _ \text{SH-Pt} \times \lambda _ \text{Pt} = 0.2$ nm. The spin pumping voltage is also observed in a symmetrical stacking, Al/FePt/Al with a lower overall efficiency. Thus, it is possible to tune and increase the charge current produced on Si//Al/FePt/Al by a factor of 1.3 on Si//FePt/Al and a gain of 2.4 on a single Si//FePt. Moreover, by studying bilayer systems such as Si//FePt/Pt and Si//Pt//FePt we independently could extract the individual contributions of the external inverse spin Hall effect of Pt and the self-induced inverse spin Hall effect of FePt. Notably, this method gives consistent values of charge currents produced due to only self-induced inverse spin Hall effect in FePt layers. These results advance our understanding of spin-to-charge interconversion mechanisms in composite thin films and pave the way for the development of next-generation spintronics devices based on self-torque.
\end{abstract}

\maketitle 
\newpage

\section{Introduction}
The interconversion between charge and spin currents through the spin Hall effect (SHE) and the inverse spin Hall effect (ISHE) has been generally studied in non-magnetic heavy metals (HM) thin films with strong spin-orbit coupling (SOC) \cite{Sinova2015,Ando2011,Du2015}. Earlier experiments consisting of the generation of a pure spin current in a ferromagnet/HM (FM/HM) interface via thermal gradients \cite{Uchida2010,Seki2015,Tian2016,Anadon2022,Anadon2021}, optical pulses \cite{Jungfleish2018, Seifert2018} or spin pumping - ferromagnetic resonance (SP-FMR) \cite{Sanchez2013,Gomez2014,Conca2016} experiments required HM layers to convert the spin current into a charge current via the ISHE or the inverse Rashba-Edelstein effect. However, new trends show that a single FM layer can also play the role of a good spin current generator/detector. From this perspective, several strategies have been explored, including the use of ferrimagnets and single FM materials with substantial spin-orbit coupling. Experiments performed by Miao \textit{et al}. \cite{Miao2013} showed that permalloy (Py) can be used to detect a spin current in a similar way as conventional HM with strong spin-orbit interaction such as Pt, Ta, or W. In that study, the longitudinal spin Seebeck effect (SSE) was used to inject a spin current from a YIG film into a Py layer, which is converted to a charge current via ISHE. 
Tsukahara \textit{et al.} \cite{Tsukahara2014} and Gladii \textit{et al.} \cite{Gladii2019} also showed that a single Py layer can convert a spin current into a charge current by self-induced ISHE SP-FMR. This has also been shown in iron and cobalt single layers \cite{Kanagawa2017}. On the other hand, self-generated spin accumulation interacting with its own magnetization, self-torque, has been also reported in single ferromagnetic layers such as CoFeB \cite{PhysRevApplied.17.064052}, or a quasi-isolated ferrimagnetic such as GdFeCo \cite{Damas2022,CespedesBerrocal2021}. FePt alloys are also good candidates to explore the spin-to-charge interconversion with the added advantage that it is a material used in hard disks to store information. The strong SOC and perpendicular magnetic anisotropy make FePt alloys good candidates for the development of new spintronic devices. In this frame, self spin-orbit torques in crystalline L1$_{0}$ FePt single layers \cite{Liu2020, Tang2020} and inverse spin Hall effect in disordered FePt \cite{Seki2015} have demonstrated efficient conversion of spin currents into charge currents and vice-versa. It has been shown that the disorder in perpendicularly magnetized FePt films also plays a key role in the generation of spin currents and on the spin-orbit torque (SOT) efficiency \cite{Zheng2020}. An efficient torque that can be exerted by epitaxial FePt on another magnetic layer, FePt/Cu/NiFe, has also been reported \cite{Seki2019}. In this work a spin anomalous Hall effect efficiency of 0.25 is reported, although the contribution of the Oersted field may also play an important role \cite{Bello2022}.

The spin-to-charge interconversion processes in single ferromagnetic layers have been explored by several groups \cite{Tsukahara2014, Gladii2019, Kanagawa2017}. Tsukahara \textit{et al.} \cite{Tsukahara2014} have attributed the symmetric contribution to self-induced ISHE to the spin currents generated by the spatial dependence of damping, which is stronger in the region near the substrate. This spin current propagates towards the substrate and is converted to charge current by ISHE \cite{Miao2013}. Similarly, the composition gradient of FePt shows naturally a slightly richer Pt region near the top interface (opposite to the substrate) and it is at the origin of self-torque \cite{Liu2020, Tang2020}. However, the quantification of the efficiency of such interconversion and the value of a characteristic length for such conversion, namely the spin diffusion length, was not reported.

In this study, we conducted a comprehensive investigation of the reciprocal effect, the spin-to-charge current conversion characteristics of chemically disordered ferromagnetic FePt thin films by SP-FMR. After structural, magnetotransport, and static magnetic characterization, we carry out our main investigation of self-induced ISHE by FePt thickness dependence and further by analyzing multilayers and bilayers with different stacking orders. These results will help the understanding of self-induced ISHE mechanisms in magnetic thin films devices, and establish a reliable protocol to characterize such relevant spintronic parameters in a single ferromagnetic layer. Namely, its spin Hall angle and its spin diffusion length.

\section{Experimental details}
Chemically disordered FePt thin films (fcc A1-phase, in which atomic sites are randomly occupied either by Fe or Pt) were grown on naturally oxidized Si(100) at room temperature by dc magnetron sputtering with a base pressure around 1 $\times$ 10$^{-6}$ Torr. A 3.8 cm-diameter target with an Fe:Pt atomic ratio of approximately 45:55 was used \cite{Salica2010}. The films were deposited using an Ar pressure of 2.8 mTorr and 20 W of sputtering power. We have fabricated 8 samples with thicknesses $t_\text{FePt}$ varying from 3 nm to 25 nm to avoid the presence of stripe-like magnetic domains generally observed for $t_\text{FePt}$ above 30 nm \cite {Burgos2011}. The layer thicknesses were controlled using the sputter-deposition rates obtained prior to the sample preparation. No capping layers were used due to the relatively good oxidation resistance of FePt \cite{Burgos2011}. The different stacks used as control systems were also grown by the sputtering technique.  
Transmission electron microscopy (TEM) analysis were carried out with a JEM-ARM 200F Cold FEG TEM/STEM (Scanning TEM) operating at 200 kV, coupled with a GIF Quantum 965 ER and equipped with a spherical aberration (Cs) probe and image correctors (point resolution 0.12 nm in TEM mode and 0.078 nm in STEM mode). High resolution-TEM (HR-TEM) micrographs were performed to study the atomic structure of the deposited layers. Energy dispersive spectroscopy (EDS) mapping analysis were carried out systematically on the different samples to confirm the nominal composition and thickness of the different materials. The resistance was obtained from four-point probe measurements. The angular measurements of anisotropic magnetoresistance (AMR) and planar Hall effect (PHE) were performed by mounting the samples on a vertical holder stage and applying a 200 mT magnetic field parallel to the film plane in each angle. Magnetization curves of the FePt films were measured in a LakeShore Vibrating Sample Magnetometer (VSM) model 7300 applying a magnetic field parallel to the film plane. Ferromagnetic resonance (FMR) and inverse spin Hall effect were measured at room temperature in a commercial Bruker Elexsys E500 spectrometer using a rectangular microwave cavity (mode TE$_{102}$) at a frequency of 9.70 GHz (X-band) using a modulation frequency of 100 kHz and a modulation field amplitude of 0.5 mT. The power of the microwave excitation was 126.2 mW. The samples were placed at the center of the resonance cavity to get a nodal position in order to maximize the microwave magnetic field and minimize the microwave electric field. In this configuration our measurements minimize spin rectification effects (SRE) due to the microwave electric field. A static magnetic field was applied parallel to the film plane. 
For the ISHE experiments, samples were cut in a rectangular shape of $\approx3\times1$ mm$^2$ and two electrodes were attached with silver paste to each end of the samples. Additional details about the ISHE measurements can be found in our previous works \cite{Gomez2014, Gomez2016}. 

FePt films were further characterized by variable-frequency FMR from 6 GHz to 23 GHz at room temperature in a probe station with an in-plane field up to 0.6 T by using a microwave with an input power of 32 mW (15 dBm). This characterization was done using spin-pumping devices similar to the ones used for previous studies \cite{Arango2023, Fache2020, Gudin2023}.
The fabrication of these devices involved conventional UV lithography techniques. Initially, the entire device stack was patterned, and subsequent ion milling was carried out while precisely controlling the milled thickness using an ion mass spectrometer with a 4-wave IBE14L01-FA system.
In a subsequent step, we grew a 100 nm thick insulating SiO$_2$ layer by radio-frequency (RF) sputtering. This was achieved by using a Si target and a combination of Ar$^+$ and O$^{2-}$ plasma in a Kenositec KS400HR PVD system.
For the final lithography step, we patterned the contacts and performed evaporation using a PLASSYS MEB400S evaporator. The active bar had dimensions of 10 $\times$ 600 $\mu$m$^{2}$. Due to the narrow width of the bar, we anticipated minimal artifacts arising from rectification effects in the spin pumping signal \cite{Sanchez2013,martin2022suppression}.
Throughout our experiments, the device geometry remained consistent, encompassing the SiO$_2$ insulator thickness, the coplanar waveguide dimensions, and the lateral dimensions of the ion-milled samples. This uniformity allowed for a reliable comparison of the spin pumping voltage across all the devices presented in this study.
While the sample stack extended beyond the coplanar waveguide, resulting in a non-uniform RF field along the magnetic wire, this inhomogeneity did not impact the shape of the measured spin pumping signal or the primary findings presented in this manuscript, as it remained consistent across all devices.

\section{Characterization}
\subsection{Structural characterization and elemental composition}
\begin{figure}[ht]
    \centering
    \subfigure{\includegraphics[width=0.45\textwidth]{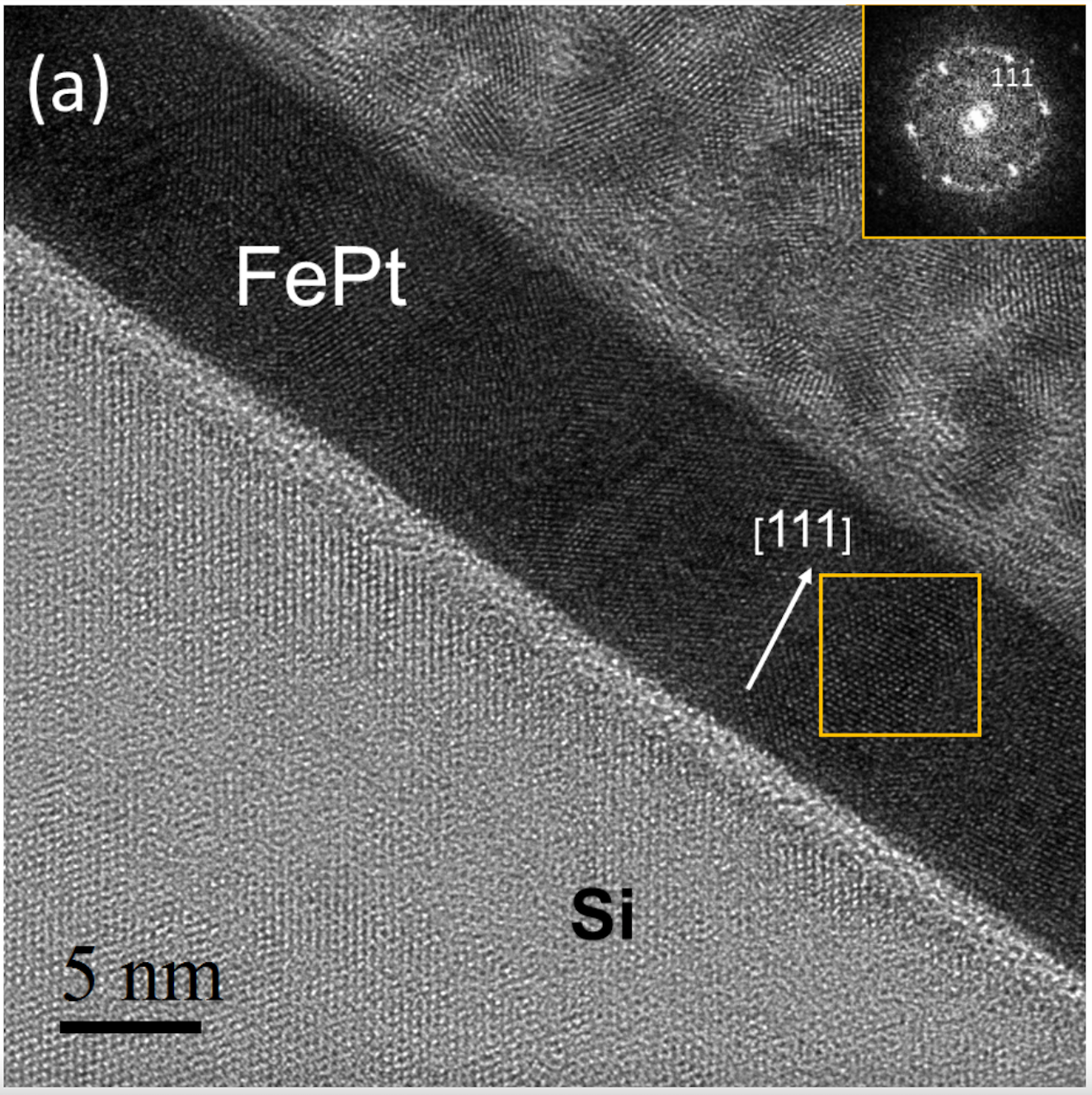}}
    \subfigure{\includegraphics[width=0.45\textwidth]{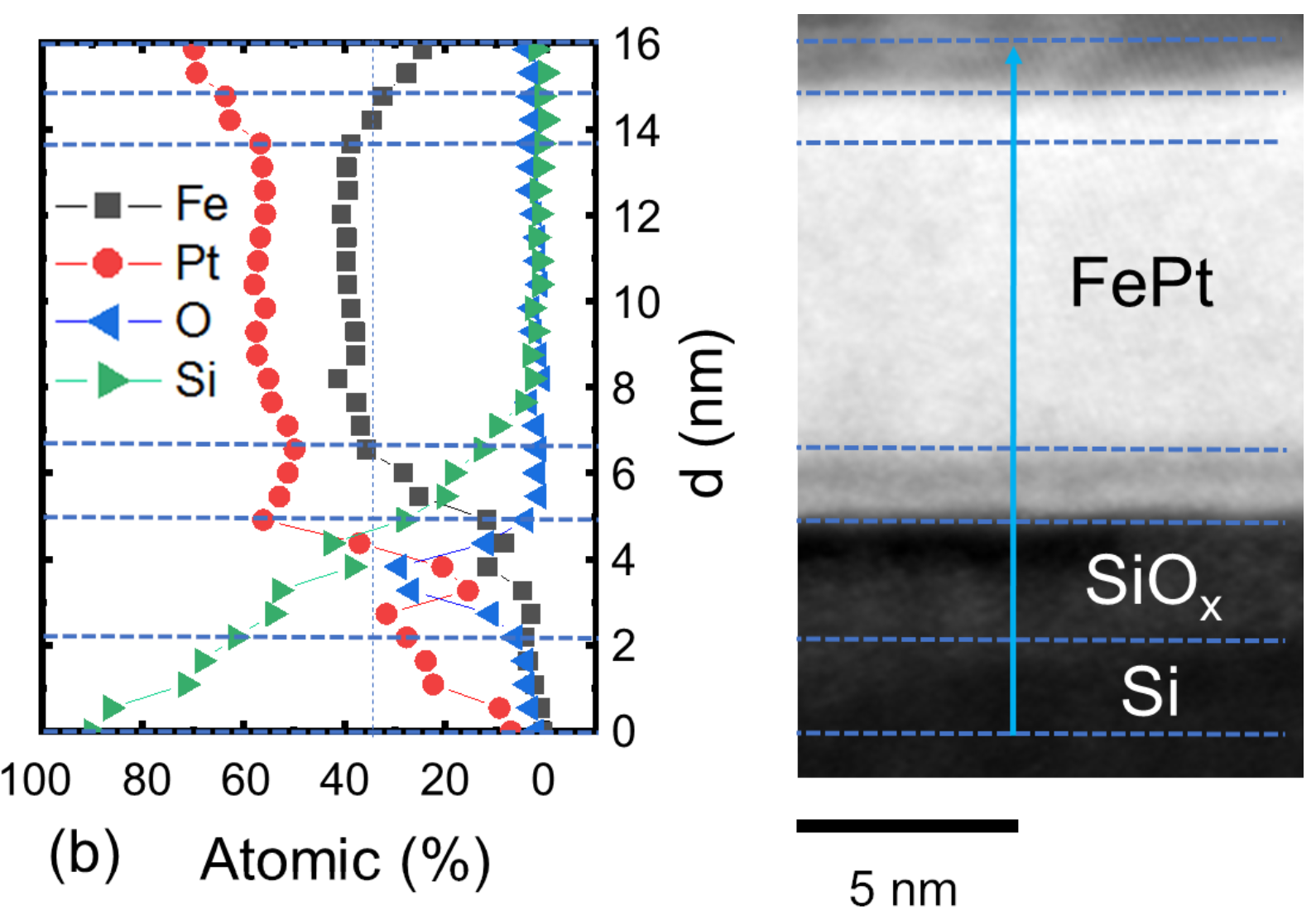}}
    \caption{(a) Cross-sectional HRTEM micrograph of a 10 nm thick FePt film deposited on a (100) silicon substrate with a native oxide layer. The FFT pattern shows the FePt $[111]$ texture. (b) EDS analysis across the dotted line shows that the elemental average composition is Fe$_{42}$Pt$_{58}$. The horizontal dashed lines indicate the transition between regions of different atomic composition. Both top and bottom interfaces display a gradient in composition in a region of approximately 1 nm.}
    \label{fig:TEM01}
\end{figure}

Fig. \ref{fig:TEM01}(a) shows a cross-sectional HRTEM micrograph of a 10 nm thick FePt single layer deposited on a (100) silicon substrate. We can observe relatively sharp interfaces and a good agreement between the nominal and the real thickness. The fast Fourier transform (FFT) pattern shows that the FePt layer is locally [111] textured, consistent with previous studies \cite{Salica2010}. Grains with different orientations are also observed in the HRTEM micrograph. Additionally, EDS analysis shows a composition Fe$_{42}$Pt$_{58}$ which is close to the nominal value 45:55, Fig. \ref{fig:TEM01}(b). We observe a slightly vertical composition gradient in the FePt samples around the center region which is strongly enhanced at the SiO$_x$-FePt and FePt-ambient interfaces. It seems that there is a richer Pt region at the top interface consistent with the previous studies of epitaxial L1$_0$ FePt on SrTiO$_{3}$/TiN(5 nm)  \cite{Liu2020} or MgO \cite{Tang2020} substrates.

\subsection{Anisotropic magnetoresistance and planar Hall effect}
\begin{figure}[H]
    \centering
    \subfigure{\includegraphics[width=0.48\textwidth]{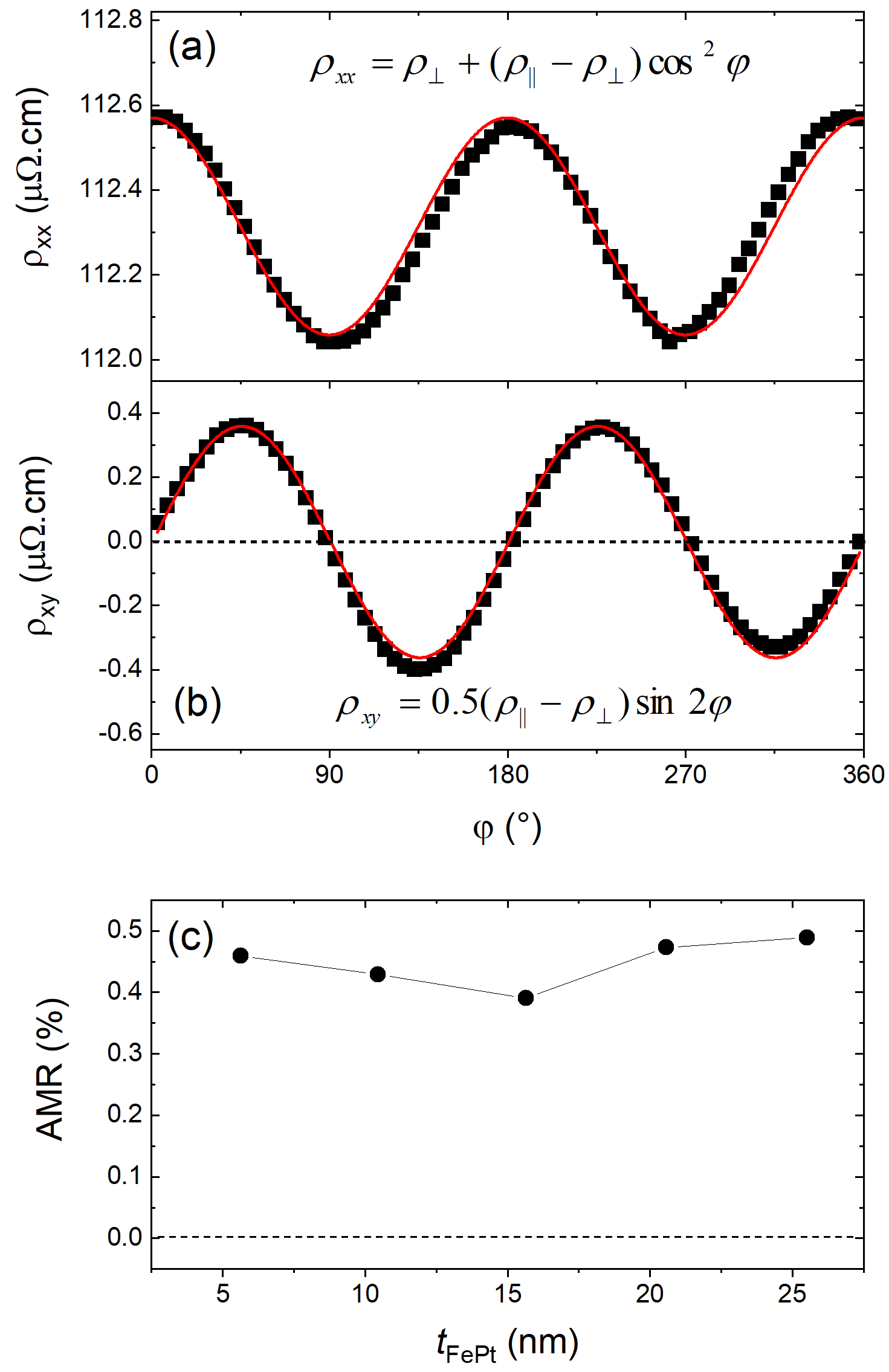}}
    \caption{(a) - (b) Angular dependence of $\rho _ \text{xx}$ and $\rho _ \text{xy}$ for a 5 nm thick FePt sample measured with a 0.2 T magnetic field applied parallel to the film plane. Solid lines are fittings with the equations displayed in the figure. (c) Thickness dependence of the AMR.}
    \label{fig:R01}
\end{figure}

  Fig. \ref{fig:R01}(a) and (b) show the angular dependence of AMR and PHE for a 5 nm thick FePt sample when a 0.2 T magnetic field is applied parallel to the film plane. For all the samples the estimated AMR$=\frac{\Delta \rho}{\rho}$ ($\Delta \rho = \rho _{\|} -\rho _{\perp}$) is less than $0.50$ \% ranging from $0.39$ \% to $0.49$ \% as shown in Fig. \ref{fig:R01}(c), which is lower than the AMR obtained, for example, in permalloy films ($\approx$ 1 \%) \cite{Rijks1997, Yamamoto1996}. As FePt films exhibit AMR and PHE, a SOC is expected in this alloy. The resistance of FePt films increases linearly with the inverse of the film thickness from 5 nm to 25 nm (see Fig. S.1 in Supplemental material), which allows the extraction of a constant resistivity value of $(110 \: \pm \: 8 )\: \mu \Omega.\text{cm}$. This value is comparable to that obtained by Hao \textit{et al.} \cite{Hao2017} for an Fe$_{47}$Pt$_{53}$ thin film alloy measured at room temperature.

\subsection{Static magnetization characterization}
\begin{figure}[H]
    \centering
    \includegraphics[width=0.48\textwidth]{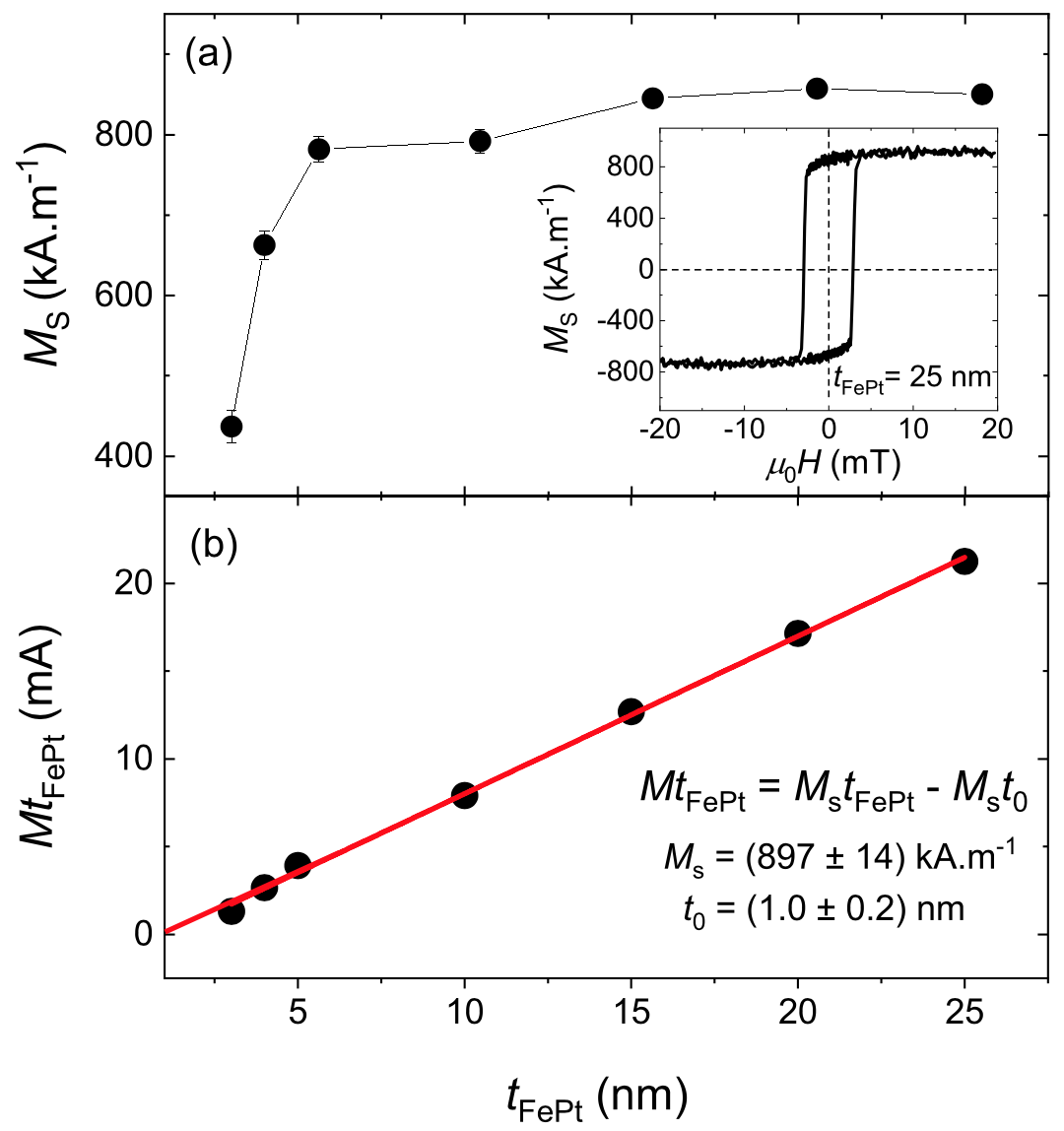}
    \caption{(a) Saturation magnetization as a function of film thickness for Si//FePt($t_\text{FePt}$). $M_\text{S}$, determined from VSM measurements, increases as the FePt thickness increases, suggesting the presence of a magnetically dead layer. The inset shows the hysteresis loop for a 25 nm thick FePt layer measured at room temperature with the magnetic field applied parallel to the film plane. The coercivity field is $2.0$ mT indicating a relatively soft ferromagnetic behavior. (b) $Mt_\text{FePt}$ $vs$ $t_\text{FePt}$ data for Si//FePt($t_\text{FePt}$). The solid line is a linear fit consistent with the presence of a magnetically dead layer of $(1.0 \pm 0.2)$ nm and a magnetic saturation of $M_S = (897 \pm 14) \: \text{kA.m}^{-1}$.}
    \label{fig:VSM01}
\end{figure}

Fig. \ref{fig:VSM01}(a) displays the dependence of the saturation magnetization ($M_S$), after VSM measurements, as a function of the thickness ($t_\text{FePt}$). $M_S$ increases from +$437 \: \; \text{kA.m}^{-1}$ at $3 \: \; \text{nm} $ to reach a relatively constant value ($850 \: \; \text{kA.m}^{-1}$) above $15  \: \text{nm}$. Fig. \ref{fig:VSM01}(b) shows magnetization per unit area, $Mt_\text{FePt}$, as a function of film thickness. A linear fit to these data suggests the existence of a magnetically dead layer of approximately $(1.0 \pm 0.2)  \: \text{nm}$ and from the slope we obtained the saturation magnetization $M_S = (897 \pm 14) \: \text{kA.m}^{-1}$. The $M_S$ value is similar to previous studies \cite{Hao2017, Salica2010}, but smaller than the bulk value of the stoichiometric compound Fe$_{50}$Pt$_{50}$ ($M_S (300\: \text{K}) = 1140 \: \text{kA.m}^{-1}$). As shown in the inset in Fig. \ref{fig:VSM01} (a),  $M$ is easily saturated with a relatively low magnetic field ($\mu _0 H \leq 5 \: \text{mT}$) which is typical of the magnetically soft A1 phase. 

\subsection{Broadband ferromagnetic resonance}
From FMR measurements we can determine different magnetic parameters such as anisotropy constants, gyromagnetic ratio and damping constant. FMR spectra carried out in a resonance cavity have been collected for all the samples in X-band (9.70 GHz) applying a magnetic field parallel and perpendicular to the film plane as shown in Fig. \ref{fig:SPschematics}(a).

We have observed a single resonance absorption (see Fig. S.2 in Supplemental material) that can be related to the uniform precession mode of the magnetization vector. From the in-plane angular dependence of the resonance field (not shown), a small easy in-plane uniaxial anisotropy field is estimated ($\approx$ 5 mT). This anisotropy field is considerably smaller than the shape contribution and for simplicity we will not consider this contribution. 

\begin{figure}[H]
    \centering
    \subfigure{\includegraphics[width=0.24\textwidth]{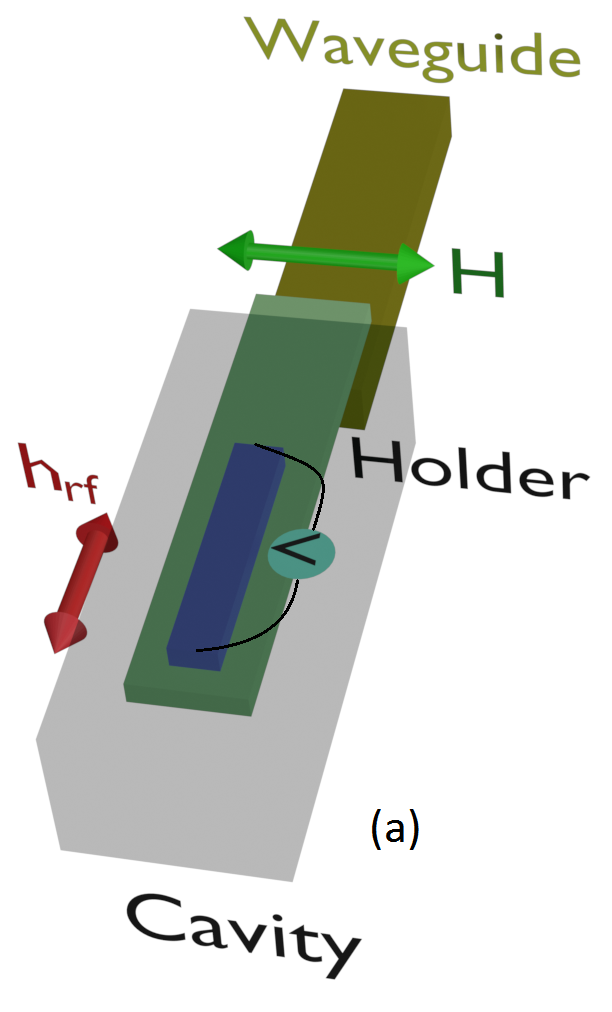}}
    \subfigure{\includegraphics[width=0.45\textwidth]{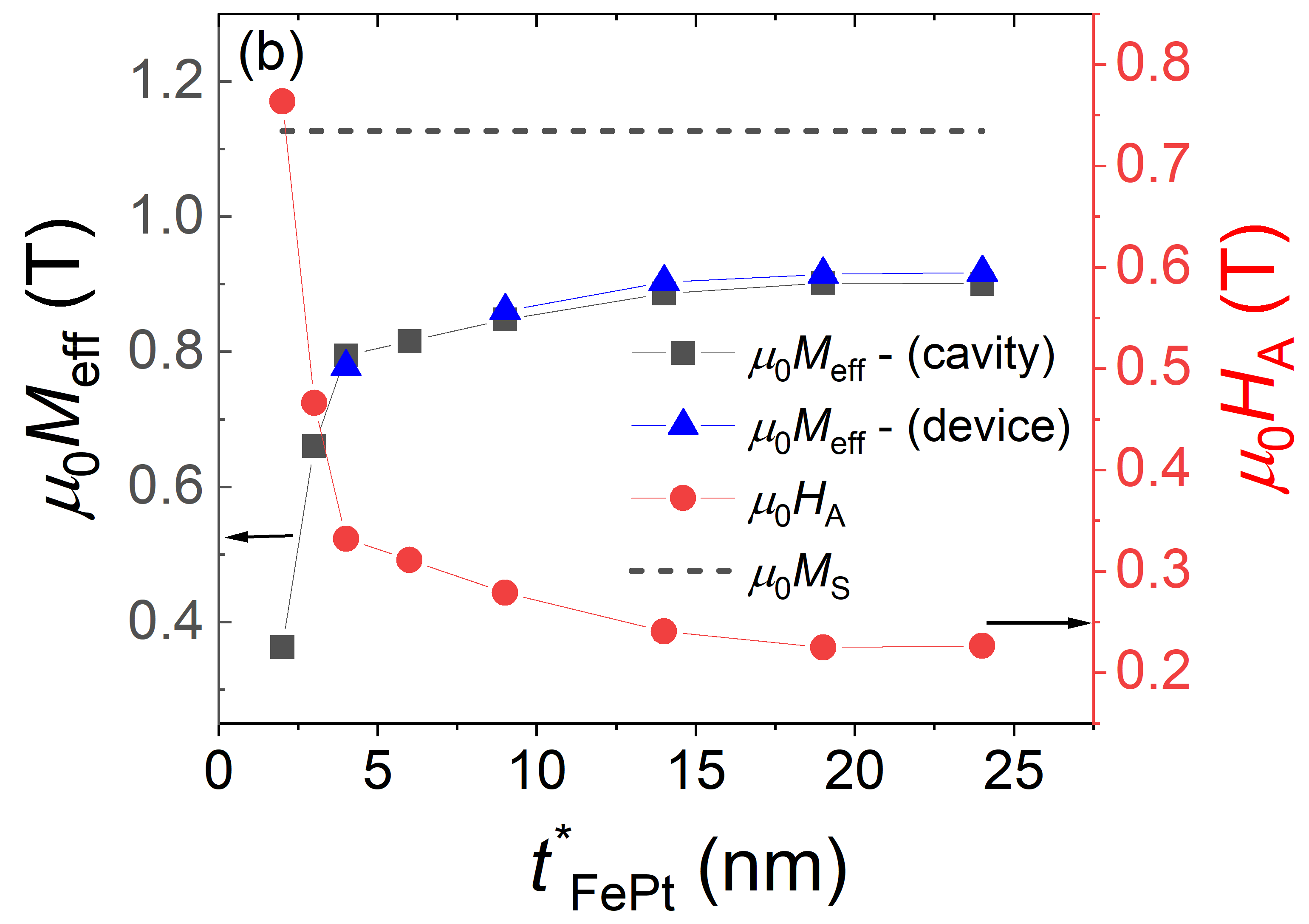}}
    \caption{(a) The sample placement inside the resonant cavity for FMR and ISHE measurements. In this scheme the DC magnetic field is applied parallel to the film plane. The microwave field $h_\text{rf}$ with the external field $H$ induce magnetization (\textit{M}) precession in the ferromagnetic layer. (b) Effective magnetization ($M_\text{eff}$) and perpendicular out-of-plane anisotropy field ($H_A$) as a function of effective film thickness $t^*_\text{FePt}$ ($t^{*}_\text{FePt}=t_\text{FePt}-t_{0}$), obtained after FMR measurements with $H$ applied parallel and perpendicular to the film plane using Eq. \ref{ec:campoefectivo2} and Eq. \ref{ec:anisotropia}, respectively. We also display $M_\text{eff}$ obtained from broadband $f$-dependance of ISHE-SPFMR measurements (blue triangles). The dashed line indicates the effective anisotropy field by considering only the shape contribution with $M_S = 897 \: \text{kA.m}^{-1}$.}
    \label{fig:SPschematics}
\end{figure}

 Then, the effective magnetization  $M_\text{eff}$ can be calculated using the FMR signal when the magnetic is applied parallel and perpendicular to the film plane (see Fig. S.2 in Supplemental material) and  the well-known Kittel formula \cite{Kittel1948}:
\begin{equation*}
      \frac{2 \pi f}{ \gamma} = \mu _0 (H_{\perp}-M_\text{eff}),
\end{equation*}
\begin{equation}
    \frac{2 \pi f}{ \gamma} = \mu _0 \sqrt{ H_{||} (H_{||}+M_\text{eff})},
    \label{ec:campoefectivo}
\end{equation}
where $H_{\perp}$ and $H_{||}$ refer to the resonance fields when the static magnetic field is applied perpendicular or parallel to the film plane, respectively.
 $f$ is the microwave excitation frequency, $\gamma$ is the gyromagnetic ratio and $M_\text{eff}$ is the effective magnetization. Rearranging Eq. \ref{ec:campoefectivo}, we get:
 \begin{equation}
    M_\text{eff} = \frac{(2 H_{\perp} + H_{||})- \sqrt{(2 H_{\perp} + H_{||})^2-4(H_{\perp} ^2- H_{||} ^2) }}{2} ,
    \label{ec:campoefectivo2}
\end{equation}
 $M_\text{eff}$ is obtained using Eq. \ref{ec:campoefectivo2}, which is shown in Fig. \ref{fig:SPschematics}(b) where $M_\text{eff}$ increases as the thickness increases. Furthermore, $M_\text{eff}$ is lower than the obtained from magnetization data $M_S$ ($M_S = 897 \: \text{kA.m}^{-1}$). This suggests the existence of a perpendicular out of plane anisotropy $H_{A}$. $M_\text{eff}$ depends on the saturation magnetization and the perpendicular out of plane anisotropy as:
\begin{equation}
     H_{A}= M_S - M_\text{eff}, 
    \label{ec:anisotropia}
\end{equation}
where $H_{A}$ is a combination of stress induced anisotropy and interfacial anisotropy. Álvarez \textit{et al.} \cite{Alvarez2016} showed that stress effects are the predominant contribution to $\mu _0 H_{A}$ for FePt films on naturally oxidized Si (100) substrates. Using Eq. \ref{ec:anisotropia}, we obtained $H_A$ for all the samples and found that it decreases from 764 mT for 3 nm FePt thick as the film thickness increases until it reaches a relatively constant value $\mu _0 H_{A} \sim$ $225$ mT beyond $20$ nm, as shown in Fig. \ref{fig:SPschematics}(b). The perpendicular anisotropy $K_{\perp}$ ($K_{\perp}= \mu _0 H_{A} M_\text{S}/2$) varies from $3.43 \times 10^5$ J/m$^3$ for 3 nm to $1.01 \times 10^5$ J/m$^3$ for 20 nm FePt thick which are consistent with the value obtained by Álvarez \textit{et al.} \cite{Alvarez2016}. The anisotropy field follows roughly a $1/t^{*}_\text{FePt}$ law, suggesting that aside from strain \cite{Alvarez2016}, surface anisotropy plays a role, especially in thinner films. Moreover, using Eq. \ref{ec:campoefectivo} for in-plane and out-of-pane measurements we can also get $\gamma = g \mu_B / \hbar$, which gives an average $g$-value of $2.08 \pm 0.03$, similar to previous reports \cite{Vasquez2008}. 

\begin{figure}[H]
    \centering
    \subfigure{\includegraphics[width=0.48\textwidth]{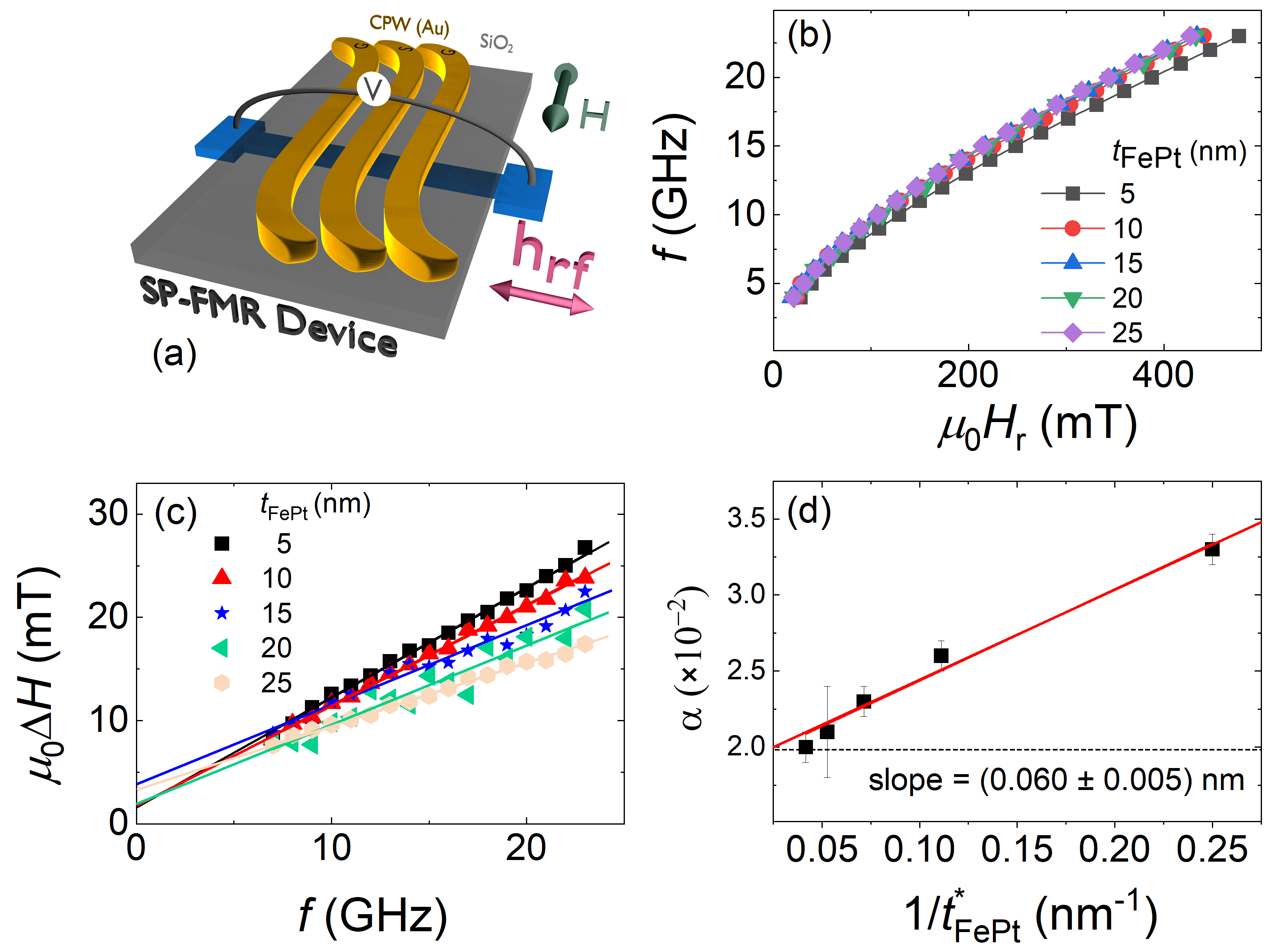}}
    \caption{(a) Schematics of a device for spin pumping ferromagnetic resonance measurements. A dc magnetic field (\textit{H}) is applied in-plane while an RF magnetic field is generated by injecting an RF electric current into the coplanar waveguide (highlighted in yellow). (b) Dispersion relationship $f$ vs $H_r$ ($H_r$: resonance field). For a fixed frequency, $H_r$ decreases as the thickness increases, suggesting that $M_\text{eff}$ increases. Fitting the data, $M_\text{eff}$ is determined for each thickness, with values similar to those obtained in FMR measurements in the resonant cavity as shown in Fig. \ref{fig:SPschematics}(b). (c) Frequency dependence of the linewidth (symbols) with the corresponding linear fits (solid lines) using Eq. \ref{ec:deltaH01}. (d) Damping constant $\alpha$ vs inverse effective film thickness $t^{*}_\text{FePt}$ (without the dead layer $t_0=$ 1.0 nm). The continuous red line is the fitting line using Eq. \ref{ec:spinn01}. It yields $\alpha _0 = 0.018$ and $g^{\uparrow \downarrow} _\text{eff}= (3.5 \pm 0.3) \times 10^{19} \text{m}^{-2}$.}
    \label{fig:damping01}
\end{figure}

In order to obtain the magnetic damping constant we have performed broadband dependence measurements of ISHE spin pumping voltage in patterned devices as shown in Fig. \ref{fig:damping01}(a) for 5 nm, 10 nm, 15 nm, 20 nm and 25 nm thick FePt layers. The frequency dependence of the linewidth can be approximately described as \cite{Gomez2014, Gomez2004}:
\begin{equation}
   \Delta H= \Delta H_0 + \frac{2 \pi \alpha}{\mu _0  \gamma} f,
    \label{ec:deltaH01}
\end{equation}
where $\alpha$ is the magnetic damping and $\Delta H_0$ is a frequency-independent term due to inhomogeneities arising from the distribution of grain sizes, shapes, anisotropies, and other scattering mechanisms.  From the $f$-dependent results we have obtained the data displayed in the Fig. \ref{fig:damping01}(b) to calculate $M_\text{eff}$ (shown in Fig. \ref{fig:SPschematics}(b)) and to get $\alpha$, Fig. \ref{fig:damping01}(c). There is a linear dependence of the linewidth with frequency, which suggests that nonlinear contributions, such as two-magnon scattering processes can be neglected. Using Eq. \ref{ec:deltaH01} we calculated the intrinsic damping for different thicknesses, all of which are comparable to the values obtained in a previous study \cite{Alvarez2013}. The dependence of $\alpha$ with $t^*_\text{FePt}$ ($t^{*}_\text{FePt}=t_\text{FePt}-t_{0}$) is shown in the Fig. \ref{fig:damping01}(d). A linear behavior of $\alpha$ vs $1/t^*_\text{FePt}$ suggests the presence of an additional relaxation mechanism of interfacial origin, which will be discussed in the next section.

\section{Results}
\subsection{Spin current measurements}
\begin{figure}[H]
    \centering
     \subfigure{\includegraphics[width=0.45\textwidth]{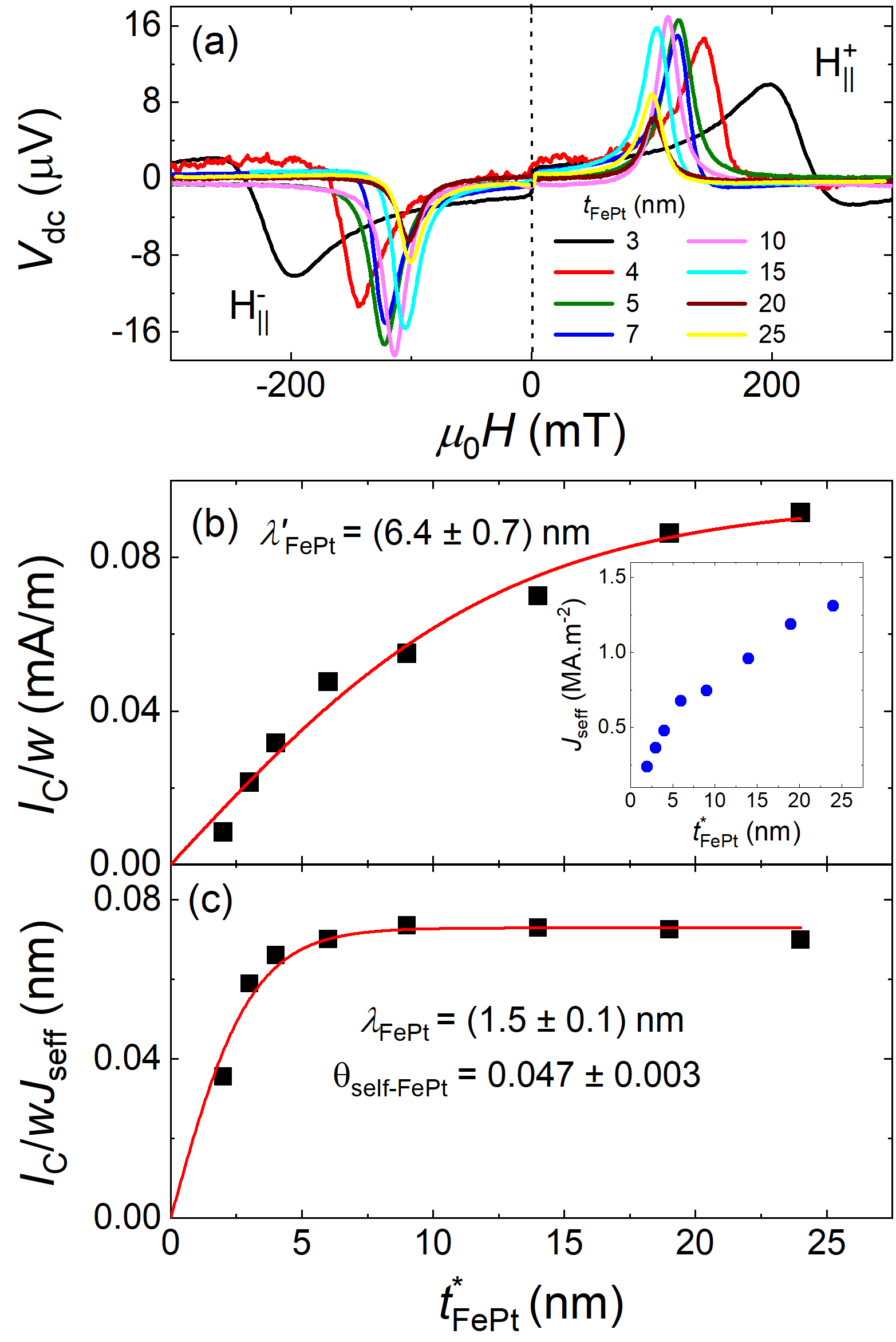}} \\
     
    \caption{(a) DC spin pumping voltages measured on FePt slabs in a resonant X-band cavity. Due to the effect of the perpendicular anisotropy the resonance field is shifted for different FePt thicknesses. (b) Charge current, normalized by the sample width ($I_\text{C}/w$) $vs$ $t^*_\text{FePt}$. The red curve is a $\tanh(t^*_ \text{FePt}/2\lambda _ \text{FePt})$ fit which gives a misleading characteristic length of 7.1 nm due to the variation of $M_\text{eff}$. This $M_\text{eff}$'s variation changes the effective spin density current ($J_{\text{seff}}$) as displayed in the inset. (c) Similar dependence as in (b) but now the charge current produced is also normalized by the injected spin current density $I_\text{C}/(w J_\text{seff})$. From the fitting, we estimated a more realistic value of $\lambda _ \text{FePt} = 1.5$ nm, and consequently of self-induced $\theta _ \text{SH-FePt} = 0.047$. 
    }
    \label{fig:YIGFePt01}
\end{figure}

In this section we will present our study of the thickness dependence of FePt. To do so, we first recall the spin pumping voltage and its analysis. We measured the rectified voltage, $V_\text{dc}$, under microwave excitation as a function of the magnetic field applied parallel to the film plane for all the samples. The voltage signal, $V_\text{dc}$ can be separated into a symmetric and an antisymmetric component according to the equation \cite{Saitoh2006}:
 \begin{equation}
 \begin{split}  
 \centering
V_\text{dc}  = & \: V_\text{S} \frac{\Delta H ^2}{(H-H_\text{r})^2+\Delta H^2}+V_\text{A} \frac{\Delta H  (H-H_\text{r})}{(H-H_\text{r})^2+\Delta H^2} \\  
+  V_\text{offset},
 \label{ec:promedio}
 \end{split}
\end{equation}
where $H_\text{r}$ is the resonance field and $\Delta H$ is the linewidth (half width at half maximum). $V_\text{S}$ and $V_\text{A}$ denote the amplitudes of the symmetric and antisymmetric components of $V_\text{dc}$, respectively. The first term can contain contributions of anomalous Nernst effect (ANE), AMR, PHE and ISHE, while the second term corresponds to the anomalous Hall effect (AHE) as long as the phase difference between the current inside the sample and the magnetization is $\pi/2$ \cite{Iguchi2017,Arango2023, Azevedo2011,Rojas-Sanchez2013}. 
Finally, the third term corresponds to a nonresonant offset voltage.  Fig. \ref{fig:YIGFePt01}(a) shows the $V_\text{dc}$ voltage measured for different FePt single layers from 3 nm to 25 nm. As the thickness increases, the resonance field decreases indicating that $M_\text{eff}$ increases as shown in Fig. \ref{fig:SPschematics} (b) due to the effect of the perpendicular out-of-plane anisotropy. We can also observe an almost symmetric signal which is odd under field inversion, see Fig. \ref{fig:YIGFePt01}(a). From the possible contributions, ANE originates from microwave-induced temperature gradients through the sample under the FMR condition. This temperature gradient is perpendicular to the film plane, which generates a charge current density in the FePt film giving a lateral dc electromotive force perpendicular to the magnetization. In our experiment, we measured $V_\text{dc}$ in the sweeping times ranging from 3 to 10 minutes, and we found that $V_\text{dc}$ does not depend on the sweep time, which suggests that ANE can be ruled out. Additionally, during the $V_\text{dc}$ voltage measurements for negative ($H_ {||} ^ {-}$) and positive ($H_ {||} ^ {+}$) values of the magnetic field, the jump of the voltages observed at $H$= 0, for all the samples, were smaller than the amplitude of the $V_\text{dc}$ voltage, see Fig. \ref{fig:YIGFePt01}(a), which gives an additional indication that the thermovoltages associated to ANE can be neglected \cite{Iguchi2017,Arango2023}. SRE originated from the nonlinear coupling of the microwave-induced charge current $I(t)$ and the field-dependent resistance ($R(H)$) \cite{Azevedo2011,Rojas-Sanchez2013}, can also be ruled out. These effects might have one antisymmetric and/or symmetric contributions corresponding to AMR, PHE or/and AHE. $V_\text{AMR}$ can be discarded because it is an even function of $H$, which is contrary to the measured $V_\text{S}$ as shown in Fig. \ref{fig:YIGFePt01}(a).
Thus, $V_\text{AMR} \thicksim 0$. 
Using Eq. \ref{ec:promedio}, the dc voltage can be separated in its symmetric and antisymmetric components as shown in Fig. S.7 in Supplemental material for a 10 nm thick FePt layer, where the symmetric component is much larger than the antisymmetric one and $V_\text{A} \thicksim 0$, so that the AHE/PHE which contribute to the antisymmetric component are small, indicating that the electric field induced by the microwave on the sample is minimum. This condition suggests that SRE are small, i.e., the voltages corresponding to either PHE or AHE are both negligible \cite{Azevedo2011,Rojas-Sanchez2013} for $t_\text{FePt} \geq 5$ nm when $H$ is applied parallel to the film plane. Furthermore, following the model developed in Ref. \cite{Rojas-Sanchez2013} for the OOP angular dependence in a resonant cavity, we have made a detailed analysis and verified that the rectifying contributions to the voltage are minimal when $H$ is parallel to the sample plane. See Supplemental material for details. However, for the case of 3 nm and 4 nm, there is a sizable $V_\text{A}$ contribution but $V_\text{S}$ still remains odd under the applied magnetic field. We will show that, consistently with $V_\text{S}$ originating in the ISHE, the charge current produced obeys the ISHE physics of a FM/HM bilayer, $\tanh (t_\text{HM}/2\lambda_\text{HM})$. On the other hand, when SRE are dominant the thickness dependence of $V_S$ for both FM and HM is expected to be inversely proportional to the HM thickness or increase monotonically with the thickness of FM as shown in Ref. \cite{Iguchi2017}. We did not observe this behavior in our samples but a maximum of $V_S$ as a function of the FePt thickness. Therefore, PHE or AHE are not important contributions to the $V_\text{S}$ when $H$ is applied parallel to the film plane for all FePt thicknesses. 
Thus, we have ensured that the symmetric component is due to ISHE or self-induced ISHE and other spurious contributions to the symmetric component are negligible in our setups (resonant cavity and litographied devices). As shown in Fig. \ref{fig:YIGFePt01}(a) the main component for $V_\text{dc}$ is the symmetric component $V_\text{S}$. 

Now we will look for quantifying some FePt spintronics parameters such as the effective spin mixing conductance ($g^{\uparrow \downarrow}_\text{eff}$), self-induced spin Hall effective angle ($\theta _ \text{self-FePt}$), and the spin diffusion length ($\lambda _ \text{FePt}$). In a first rough approximation, we will consider a model of a typical FM/HM bilayer. Although the interconversion mechanism in FePt due to its composition gradient could be more complex, we can obtain reasonable values, which can be used empirically or phenomenologically for subsequent studies. Assuming that the FePt layer behaves as a "typical" FM/HM bilayer due to its composition gradient, the thickness dependence of $\alpha$ allows to extract $g^{\uparrow \downarrow}_\text{eff}$ as \cite{Heinrich2011}:
\begin{equation}
   \alpha  = \alpha _0 + \frac{g^{\uparrow \downarrow} _\text{eff} g \mu _ {\text{B}}}{4\pi M_S} \frac{1}{t^{*} _\text{FePt}},
    \label{ec:spinn01}
\end{equation}
where $g$ is the g-factor, $\mu _ {\text{B}}$ is the Bohr magneton and $t^{*}_\text{FePt}$ is the effective FePt layer thickness $t^{*}_\text{FePt}$. Fitting the data of Fig. \ref{fig:damping01}(d) with Eq. \ref{ec:spinn01}, we get $\alpha _0 = 0.018$ and $g^{\uparrow \downarrow}_\text{eff} = (3.5 \pm 0.3) \times 10^{19} \: \text{m}^{-2}$. The value of $g^{\uparrow \downarrow}_\text{eff}$ is comparable to those obtained in epitaxial Fe/Pt bilayers $(4.9 \pm 0.5) \times 10^{19} \: \text{m}^{-2}$ \cite{Conca2016} and $(1.5 \pm 0.5) \times 10^{19} \: \text{m}^{-2}$ \cite{GuillemardFePt}. 
In addition, for a "typical" FM/HM bilayer, the voltage given by ISHE in the HM layer is given by  \cite{Ando2011,Gomez2014,Fache2020,Rojas2014}:

 \begin{equation}
    V_\text{ISHE}= Rw \theta _ \text{SH}  \lambda _ \text{HM} \tanh(t_ \text{HM}/2\lambda _ \text{HM}) J _ \text{seff},
    \label{ec:voltageISHE}
\end{equation}
where $R$ is the resistance of the bilayer, $w$ is the sample width,  $\theta _ \text{SH}$ is the spin Hall angle, $\lambda _ \text{HM}$ and $t _ \text{HM}$ are the spin diffusion length and the thickness of the heavy metal layer, respectively. $J _ \text{seff}$ is the effective spin current density injected at the interface, which reads \cite{Ando2011,Gomez2014, Fache2020,Rojas2014}:
 \begin{equation}
    J _ \text{seff}= \frac{e g^{\uparrow \downarrow} _\text{eff} \gamma  h_\text{rf} ^2 [4\pi M_\text{eff}+\sqrt{(4\pi M_\text{eff})^2+(4\pi f /\gamma)^2}]}{4\pi\alpha ^2 [(4\pi M_\text{eff})^2+(4\pi f /\gamma)^2]},
    \label{ec:corrientespinISHE}
\end{equation}
where $h_\text{rf}$ is the microwave magnetic field and $e$ is the magnitude of the electron charge. From the symmetric amplitude of the voltage, $V_\text{S}$, for different FePt thicknesses and the corresponding $R$, we calculated the produced charge current $I_\text{C}=V_\text{S}/R$ and normalized it by $w$.

Fig. \ref{fig:YIGFePt01}(b) shows the $t^{*} _\text{FePt}$ dependence of the charge current normalized by the width ($I_\text{C}/w$) of the spin pumping slab experimentally measured in the X-band resonant cavity. 
These values show a characteristic $\tanh(t^*_ \text{FePt}/2\lambda _ \text{FePt})$ dependence due to ISHE after Eq. \ref{ec:voltageISHE}. The fit results in a value of 6.4 nm which is relatively high for a resistive FePt, 110 $\mu \Omega. \text{cm}$. Although in our cavity $h_\text{rf}= 0.613 \times 10^{-4}$ T and frequency 9.70 GHz, were the same in all measurements, there is another important parameter that does vary and has not been considered yet: the perpendicular anisotropy of FePt plays a relevant role and changes the macroscopic parameter $M_\text{eff}$ with FePt thickness, Eq. \ref{ec:anisotropia} and Fig. \ref{fig:SPschematics}(b). 
If we now consider all these parameters, we can calculate the effective spin density current ($J_{\text{seff}}$) using Eq. \ref{ec:corrientespinISHE} for each FePt thickness, inset Fig. \ref{fig:YIGFePt01}(b),  and thus obtain $Ic/(w J_{\text{seff}})$ whose dependence on thickness is shown in Fig. \ref{fig:YIGFePt01}(c). The result of such a fit, which is the combination of Eq. \ref{ec:voltageISHE} and \ref{ec:corrientespinISHE}, gives $\lambda _ \text{FePt}= (1.5\pm 0.1)$ nm and self-induced $\theta_ \text{self-FePt}=0.047 \pm 0.003$. These values are more realistic, a $\lambda$ of a few nm is expected for a resistive layer as in our FePt, $ \rho=(110 \: \pm \: 8 )\: \mu \Omega.\text{cm}$. 
Hao et al. have shown, using charge-charge conversion in AHE experiments, that $\lambda _ \text{FePt}= (3.9\pm 0.2 )$ nm for Fe$_{29}$Pt$_{71}$ with a bulk resistivity of $(98 \pm 1 )\: \mu \Omega.\text{cm}$  \cite{Hao2017}.
The spin diffusion length of our FePt films is comparable to a resistive heavy metal such as Ta (1.5 nm \cite{Gomez2014} or 1.8 nm \cite{Hahn2013} ), or W (1.6 nm \cite{Kim2016}). And, as expected, $\lambda _ \text{FePt}$ is lower than weaker SOC and less resistive FM materials such as Co (7.7 nm \cite{Zahnd2018}). \textcolor{red}{}
Now, we can estimate the product $\theta _ \text{self-FePt} \times \lambda _ \text{FePt} = 0.071$ nm, and the spin conductivity of FePt, $\sigma_\text{self-FePt}=\theta_\text{self-FePt}/\rho_\text{FePt}=4\times10^4$ S.m$^{-1}$. Furthermore, the method that we report will allow us to study the spin transport parameters such as the spin diffusion length of the FM layers in a much simpler way than using nanodevices \cite{Bass_2007}.

These results support our assumption of a bilayer model to quantify the spintronic parameters in our FePt single layer. Although, as previously mentioned, these values can be used empirically or phenomenologically for future studies since the interconversion mechanism in FePt would be more complex than a simple typical FM/HM bilayer. One question that arises, however, is why the bilayer model would still work in a single FePt layer. The answer would be precisely the concept of self-spin pumping and self-torque driven by a composition gradient in FM layers with strong bulk SOC. In a simplistic picture, FePt is simultaneously the FM layer and the HM layer in a typical bilayer treatment. The sign of the self-induced voltage can be tuned by changing the sign of the composition gradient. This has been shown in the direct effects on different systems such as GdFeCo, CoTb \cite{Zheng2021} and FePt \cite{Tang2020}. To verify this in the inverse effect on FePt, we will study different buffers and capping layers in the next section.

Besides, these FePt spintronic parameters are relevant for future studies and applications, either using the direct or inverse effect. For the direct effect, we find that the spin Hall conductivity on FePt is lower than that of Pt ($3.23\times10^5$ S.m$^{-1}$ \cite{Rojas2014}), but using the figure of merit for comparing direct effects, $\theta_\text{SH}/\lambda_{\text{SDL}}$ \cite{Rojas2019b}, FePt (0.028 nm$^{-1}$) would have an advantage over Pt (0.016 nm$^{-1}$). For the reciprocal effect, the product $\theta _ \text{self-FePt} \times \lambda _ \text{FePt}=0.071$ nm is comparable to that of Pt where $\theta _ \text{SH-Pt} \times \lambda _ \text{Pt}\approx0.2$ nm \cite{Rojas2014, Rojas2019b}.

Consequently, the FePt layer exhibits large self-induced ISHE as long as FePt has a non-negligible spin-orbit coupling. From the positive polarity of $V_\text{S}$ and using the equation $\vec{j} _c = \frac{2e}{\hbar}  \theta _ \text{SH} \: \: \vec{j} _s \times \hat{\sigma}$ with $\hat{j} _s$ (+z direction) and $\hat{\sigma}$ (+x direction, the same direction of the static magnetic field, $H_\text{dc}$), the sign of $\theta _ \text{SH}$ of FePt layer is positive, which is the same as that in Pt \cite{Wang2013, Azevedo2011} and opposite to that in Ta or W \cite{Wang2013}. We have independently verified this by measuring different bilayers such as NiFe/Pt, CoFeB/Ta, CoFeB/Pt, and so on. We note that this positive sign is due to the compositional gradient that forms naturally in FePt, along with the stronger gradients at the interfaces. Such a natural gradient can be tuned to change the amplitude of the interconversion signal, or even its sign as discussed above. Hence, we can define a composition gradient according to the Pt concentration, \textit{i.e.}  $J_S || \nabla X_{Pt} ||+z$, where $+z$  goes from the substrate to the air interface. While we cannot fully exclude also a self-induced Inverse Edelstein Effect (IEE) -like contribution in our FePt system, we believe that the conversion is dominated by the physics of the self-ISHE. This is, on the one hand, because we have a sputtering-grown polycrystalline disordered FePt system, and, on the other hand, the dependence on FePt thickness follows well the ISHE physics (see Fig. 6(c)). In the case of the IEE such dependence should be constant \cite{Rojas2019b,Rojas-sanchez2013_Ag-Bi, Rojas-sanchezPRL2016}.

\begin{table}[H]
    \centering
    \resizebox{8.2cm}{!} {
    \begin{tabular}{|c|c|c|c|}
    \hline
         Sample & Stack & $I_\text{C}/w$ ($\mu$A/m) & $\Delta H _\text{pp}$ (mT)\\
         \hline\hline
         A & Si//FePt (10 nm)/Pt (7 nm)  & 140 $\pm$ 2 & 11.6\\
         \hline
         B & \textbf{Si//FePt(10 nm)}& \textbf{55 $\pm$ 1} & 12.5\\
          \hline
         C & Si//FePt (10 nm)/Al (10 nm) & 30 $\pm$  1 & 11.2\\ 
         \hline
         D & Si//Al (5 nm)/FePt (10 nm)/Al (5 nm) & 23 $\pm$ 1 & 10.1\\
         \hline
         E & Si//Pt (7 nm)/FePt (10 nm) & -67 $\pm$ 1 & 15.5\\
         \hline
    \end{tabular}
    }
    \caption{Summary of the different grown stacks of multilayers and bilayers with Pt. The measured charge current produced (spin pumping voltage normalized by the sample width and resistance) and the peak-to-peak linewidth, $\Delta H _\text{pp}$, measured at the X-band resonant cavity are displayed. We have verified that the microwave field is the same in all measurements, $h_\text{rf}= 0.613 \times 10^{-4}$ T, so a comparison of the produced charge currents can be made among the different systems.
    }
    \label{tab:stacks}
\end{table}

\begin{figure}[H]
    \centering
   \subfigure{ \includegraphics[width=0.475\textwidth]{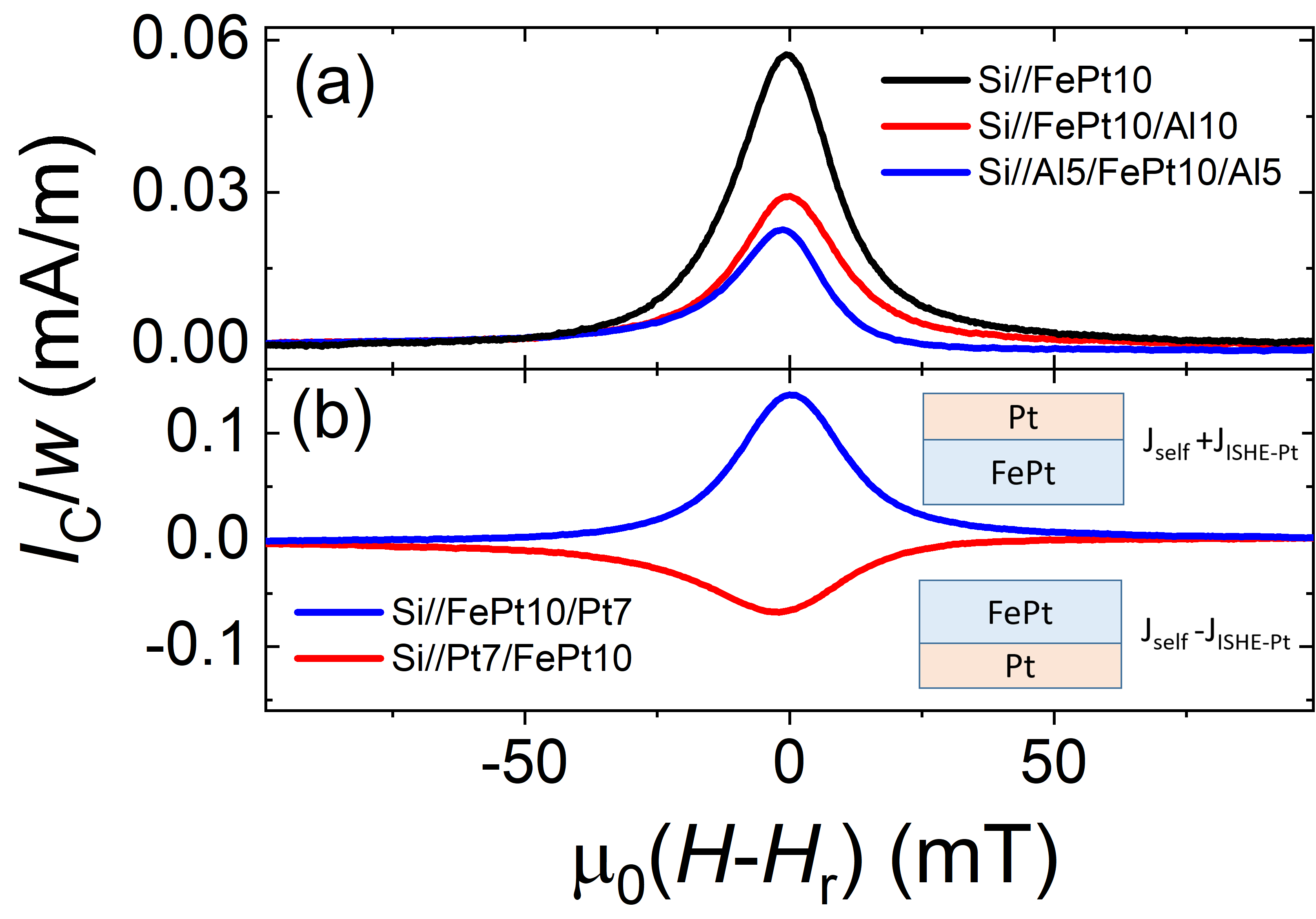}}
    \caption{(a) Raw spin pumping voltage normalized by resistance and width for different FePt slabs. (b) Same measurements for different stacking orders in the bilayers Si//FePt10/Pt7 and Si//Pt7/FePt10 used as reference samples. The different amplitudes suggest that it is necessary to take into account a self-induced ISHE in the FePt layer.}
    \label{fig:stack}
\end{figure}

In order to further understand the origin and behavior of the self-induced spin pumping voltage signal, we tuned the amplitude of our self-induced ISHE spin pumping voltage signal. To do so, we have grown different control systems whose stacks are shown in Table \ref{tab:stacks} and the $I_\text{C}/w$ signals are shown in Fig. \ref{fig:stack}(a) and (b). In the Si//FePt(10 nm) single layer, which has a strong composition gradient at both interfaces (Fig. \ref{fig:TEM01}(b)), the charge current production is +55 $\mu$A/m, sample B. Fig. \ref{fig:stack}(a) shows that the signal decreases adding an Al capping layer, and even more in an Al/FePt/Al symmetrical stack. Adding an Al layer on top of the FePt layer, sample C, the $I_\text{C} /w$ value decreases by 45 \% to a value of 30 $\mu$A/m. This can be understood by assuming that the Al capping layer reduces the composition gradient at the top interface. By depositing an Al layer on both sides of FePt to have a symmetrical stacking, we observe an additional reduction in $I_\text{C} /w$. If both interfaces of the initial Si//FePt system contribute equally, another reduction of 45\% with respect to Si//FePt/Al would be expected here as well. However, adding an Al capping and buffer layer, sample D, $I_\text{C}/w$ decreases only by 20 \% to a value of 23 $\mu$A/m compared to 30 $\mu$A/m in the asymmetric stack, sample C. In the symmetrical sample, we should have the FePt contribution with the lowest composition gradient at interfaces without native Si oxide and air. This lowest value, 23 $\mu$A/m, increases by a factor of 2.4 in the Si//FePt single layer. Our results would also indicate that in the single layer, the upper interface with the air contributes more than the lower interface with the substrate. This could occur because a compositional gradient is maintained with a Pt-rich region at the upper interface even in the Al/FePt/Al symmetrical stack \cite{Liu2020, Tang2020}.

 Now that we know the behavior of FePt single layers and multilayers, we can exploit that to confirm such results independently in a bilayer including a heavy metal with strong SOC, such as Pt. It is then expected that in a //FePt/Pt bilayer the signal, in absolute value, should be greater than that of //Pt/FePt. And that is what we observed in our experimental results Fig. \ref{fig:stack}(b). We recall that the signal changes its sign in typical FM/HM and HM/FM bilayers due to the change in the direction of spin current injection. Now that we have a strong self-induced signal in FePt, the scenario would be the following: 
 
 i) for the //FePt/Pt stacking order, the total $I_\text{C}/w= 140$ $\mu$A/m will be the addition of the extrinsic contribution of Pt, SHE of Pt, and the self-induced contribution of FePt:
 \begin{equation}
     I_\text{C}/w=I_\text{self-FePt}/w+I_\text{SH-Pt}/w= 140 \; \mu\text{A/m},
     \label{ec:add}
 \end{equation}
 ii) for the reverse stacking order, //Pt/FePt, we have the contrary, i.e a subtraction:
 \begin{equation}
     I^{*}_\text{C}/w= I^{*}_\text{self-FePt}/w-I^{*} _\text{SH-Pt}/w= -67 \; \mu\text{A/m}.
     \label{ec:subs}
 \end{equation}

 A similar method was applied but to direct effects to estimate self-torque in GdFeCo capping with Cu, Pt or Ta \cite{CespedesBerrocal2021}. As predicted, total $I_\text{C}/w$ for FePt/Pt and Pt/FePt have remarkably different amplitudes, which supports our scenario that the FePt layer contributes to the total $I_\text{C}/w$. The negative sign of $I^*_\text{self-FePt}/w-I^*_\text{SH-Pt}/w$ indicates that the extrinsic spin conversion in Pt dominates over the self-conversion in FePt. 
 
 The quality of the layers and interfaces is not expected to be the same for both samples with reverse stacking. Thus, to make a rigorous quantification would require a FePt-thickness dependence for each stacking order. This is beyond the scope of the present work. However, we can make a zero-order approximation assuming that FePt/Pt interfaces on both bilayers are identical. Then we can estimate the self-induced contribution of FePt as $I_\text{self-FePt}/w=I^{*}_\text{self-FePt}/w=(I_\text{C}/w+I^{*}_\text{C}/w)/2= 36$ $\mu$A/m. This value is lower than that of simple FePt, $i.e.$ without any protection, but higher than that of the protection with aluminum. This would happen because when depositing the Pt it unbalances the stoichiometry of the interface, producing a different composition gradient lower than the FePt/air. Indeed, we can infer also that such a compositional gradient is different in the //Pt/FePt bilayer, either from the coercivity in \textit{M}(H) loops measured in a VSM, and from the resonance field in FMR experiments. The other samples, single layer, //FePt/Pt, and other multilayers with Al have similar $H_c$ and resonance field (see Fig. S.3 and S.4 in Supplemental material).

 Using the bilayer and trilayer study, the self-induced contribution of FePt is unambiguously and independently demonstrated. The bilayer samples also indicate that the FePt layer behaves like a Pt capping layer giving a positive $I_\text{C}/w$ (spin current direction similar to a capping Pt layer) \textit{i.e.} FePt layer behaves like FM/Pt system \cite{Azevedo2011, Fache2020, Rojas2014} which is fully consistent with our results from single FePt layers. The compositional gradient is relevant for the reciprocal effect as has been shown for the direct effect \cite{Liu2020, Tang2020, CespedesBerrocal2021, Zhu2021}.

\section{Conclusions}
This study provides a comprehensive characterization of the self-induced inverse spin Hall effect in FePt. The structural analysis reveals good agreement between nominal and actual film thickness. The FePt films exhibit a [111] texture and a composition close to the nominal value while the top and bottom interfaces show a marked composition gradient with a Pt-richer region at the upper interface. Thickness dependence of magnetic characterization reveals a magnetically dead layer of ($1.0\pm0.2$) nm.

We demonstrate using spin pumping experiments that FePt can generate a spin current and convert it efficiently to a charge current by self-induced ISHE with a positive spin Hall angle, \textit{i.e.} same sign as for Pt. The results of FePt-thickness dependence allow for quantifying spintronics parameters on a single FePt layer as a "typical" FM/HM bilayer. For direct effect applications, FePt has non-negligible values: $\lambda _ \text{FePt}= (1.5 \pm 0.1)$ nm and self-induced $\theta_ \text{self-FePt}=0.047 \: \pm \: 0.003$, and 
$\theta_\text{SH}/\lambda_{\text{SDL}}=0.031$ nm$^{-1}$, for comparison the one in Pt is 0.016 nm$^{-1}$. Aditionally, for applications exploiting the reciprocal effects, 
$\theta _ \text{SH} \times \lambda _ \text{HM} = 0.071$ nm for a sputtered Si//FePt single layer is comparable to the one of Pt ($\theta _ \text{SH} \times \lambda _ \text{HM}=0.2$ nm).

In a symmetrical stacking, Si//Al(10 nm)/FePt (10 nm)/Al (10 nm) we have the lowest signal of the self-induced ISHE in FePt which can be tuned by adding a strong composition gradient at the interfaces. Thus, we increase the charge current produced by a factor of 1.3 in Si//FePt(10 nm)/Al and up to a gain of 2.4 in the single Si//FePt layer.
This result emphasizes the role of structural symmetry breaking in enabling the self-induced inverse spin Hall effect and the possible relevance of self-torque in studies using other ferromagnet/metal bilayers. These findings also reveal that FePt films can serve as efficient spin current detectors, positioning them as valuable components for future spintronic and magneto-optical applications. The methods developed in the present study can be used to reveal and quantify self-induced ISHE in other FM materials with strong SOC.

\section*{Acknowledgement}
Technical support from Rubén E. Benavides, César Pérez, Laurent Badie and Matías Guillén is greatly acknowledged. This work was partially supported by Conicet under Grant PIP 2021-00281, PIBAA 2022-2023 project MAGNETS Grant ID. 28720210100099CO, ANPCyT Grant PICT 2018-01138, ANPCyT Grant PICT 2021-00113 project DISCO and U.N. Cuyo Grant 06/C004, all from Argentina. L. Avil\'es-F\'elix thanks to Antonio Azevedo, Jinho Lim and Axel Hoffmann for fruitful discussions. We also acknowledge the support from EU-H2020-RISE project Ultra Thin Magneto Thermal Sensoring ULTIMATE-I (Grant ID. 101007825). Devices in the present study were patterned at Institut Jean Lamour's clean room facilities (MiNaLor). These facilities are partially funded by FEDER and Grand Est region through the RANGE project. This work was also supported by the French National Research Agency (ANR) through the ANR-19-CE24-0016-01 ‘Toptronic’ ANR JCJC. HRTEM characterization was performed at Institut Jean Lamour's microscope CC-3M facilities.

\bibliography{References}

\makeatletter
\noindent

\end{document}